\PassOptionsToPackage{unicode}{hyperref}
\PassOptionsToPackage{hyphens}{url}
\PassOptionsToPackage{dvipsnames,svgnames,x11names}{xcolor}
\documentclass[
  Crown]{sagej}

\usepackage{amsmath,amssymb}
\usepackage{iftex}
\ifPDFTeX
  \usepackage[T1]{fontenc}
  \usepackage[utf8]{inputenc}
  \usepackage{textcomp} 
\else 
  \usepackage{unicode-math}
  \defaultfontfeatures{Scale=MatchLowercase}
  \defaultfontfeatures[\rmfamily]{Ligatures=TeX,Scale=1}
\fi
\usepackage{lmodern}
\ifPDFTeX\else  
\fi
\IfFileExists{upquote.sty}{\usepackage{upquote}}{}
\IfFileExists{microtype.sty}{
  \usepackage[]{microtype}
  \UseMicrotypeSet[protrusion]{basicmath} 
}{}
\makeatletter
\@ifundefined{KOMAClassName}{
  \IfFileExists{parskip.sty}{%
    \usepackage{parskip}
  }{
    \setlength{\parindent}{0pt}
    \setlength{\parskip}{6pt plus 2pt minus 1pt}}
}{
  \KOMAoptions{parskip=half}}
\makeatother
\usepackage{xcolor}
\makeatletter
\ifx\paragraph\undefined\else
  \let\oldparagraph\paragraph
  \renewcommand{\paragraph}{
    \@ifstar
      \xxxParagraphStar
      \xxxParagraphNoStar
  }
  \newcommand{\xxxParagraphStar}[1]{\oldparagraph*{#1}\mbox{}}
  \newcommand{\xxxParagraphNoStar}[1]{\oldparagraph{#1}\mbox{}}
\fi
\ifx\subparagraph\undefined\else
  \let\oldsubparagraph\subparagraph
  \renewcommand{\subparagraph}{
    \@ifstar
      \xxxSubParagraphStar
      \xxxSubParagraphNoStar
  }
  \newcommand{\xxxSubParagraphStar}[1]{\oldsubparagraph*{#1}\mbox{}}
  \newcommand{\xxxSubParagraphNoStar}[1]{\oldsubparagraph{#1}\mbox{}}
\fi
\makeatother

\usepackage{longtable,booktabs,array}
\usepackage{calc} 
\usepackage{etoolbox}
\makeatletter
\patchcmd\longtable{\par}{\if@noskipsec\mbox{}\fi\par}{}{}
\makeatother
\usepackage{graphicx}
\makeatletter
\def\maxwidth{\ifdim\Gin@nat@width>\linewidth\linewidth\else\Gin@nat@width\fi}
\def\maxheight{\ifdim\Gin@nat@height>\textheight\textheight\else\Gin@nat@height\fi}
\makeatother
\setkeys{Gin}{width=\maxwidth,height=\maxheight,keepaspectratio}
\makeatletter
\def\fps@figure{htbp}
\makeatother

\makeatletter
\@ifpackageloaded{caption}{}{\usepackage{caption}}
\AtBeginDocument{%
\ifdefined\contentsname
  \renewcommand*\contentsname{Table of contents}
\else
  \newcommand\contentsname{Table of contents}
\fi
\ifdefined\listfigurename
  \renewcommand*\listfigurename{List of Figures}
\else
  \newcommand\listfigurename{List of Figures}
\fi
\ifdefined\listtablename
  \renewcommand*\listtablename{List of Tables}
\else
  \newcommand\listtablename{List of Tables}
\fi
\ifdefined\figurename
  \renewcommand*\figurename{Figure}
\else
  \newcommand\figurename{Figure}
\fi
\ifdefined\tablename
  \renewcommand*\tablename{Table}
\else
  \newcommand\tablename{Table}
\fi
}
\@ifpackageloaded{float}{}{\usepackage{float}}
\floatstyle{ruled}
\@ifundefined{c@chapter}{\newfloat{codelisting}{h}{lop}}{\newfloat{codelisting}{h}{lop}[chapter]}
\floatname{codelisting}{Listing}

\makeatother
\makeatletter
\@ifpackageloaded{caption}{}{\usepackage{caption}}
\@ifpackageloaded{subcaption}{}{\usepackage{subcaption}}
\makeatother

\usepackage{bookmark}

\hypersetup{
  pdftitle={A State-Space Approach to Modeling Tire Degradation in Formula 1 Racing},
  colorlinks=true,
  linkcolor={blue},
  filecolor={Maroon},
  citecolor={Blue},
  urlcolor={Blue},
  pdfcreator={LaTeX via pandoc}}

\title{A State-Space Approach to Modeling Tire Degradation in Formula 1
Racing}
\author{
Cole Cappello\affilnum{1} and Andrew Hoegh\affilnum{1}
}

\affiliation{\affilnum{1}Department of Mathematical Sciences, Montana State University, Bozeman, MT, USA}

\corrauth{Cole Cappello, Department of Mathematical Sciences, Montana State University, Bozeman, MT 59717, USA.}

\email{colecappello@montana.edu; andrew.hoegh@montana.edu}
\date{}

\begin{document}
\maketitle

\section{Introduction}\label{introduction}

One of the most important factors contributing to race strategy in a
Formula 1 Grand Prix is tire degradation. As tires degrade throughout
the course of a race, drivers are forced to go slower. As such, it can
be beneficial to enter the pit lane for a new set of tires. However,
drivers lose time relative to their competitors while they are waiting
for new tires to be put on. In this way, deciding to make a pitstop is a
delicate balance, and can easily affect a competitor's results. A
dramatic example of this occurred during the 2024 Italian Grand Prix, in
which Charles Leclerc of Ferrari beat Oscar Piastri of McLaren (Giles,
2024). Leclerc only stopped once for new tires, while Piastri--who was
leading the race and on track to win--made a second pitstop later in the
race that cost him victory. By stopping early for a set of ``hard''
tires and continuing till the end of the race, Ferrari was able to beat
their opponent even though their car was generally slower than the
McLaren throughout that weekend.

In each Formula 1 grand prix, there are three different dry tire
compounds for teams to choose from, along with an intermediate tire and
a full wet tire for rainy conditions. The dry tires--referred to as
``hard'', ``medium'', and ``soft''--are designed by tire manufacturer
Pirelli to degrade at different rates (Pirelli, n.d.). Tire degradation
itself is a phenomenon which occurs as a result of the extreme forces
put through the tires during a grand prix. These forces cause shearing
of the rubber from the surface of the tire and thermal degradation of
the tire carcass due to friction (Farroni et al., 2016). Softer tires
provide more grip initially and allow for faster lap times. However,
they degrade faster than harder tires, which leads to slower lap times
and a need to pit sooner. On the other hand, harder tires are not as
quick, but degrade more slowly. Therefore, a driver can typically do
more laps at a reasonable pace on a harder tire.

As tires wear, lap times tend to increase throughout a stint (a set of
laps completed on a single set of tires). Because replacing degraded
tires can yield faster overall race times, strategists must balance tire
longevity against short-term performance. Predictive models of
degradation can help answer questions such as ``How rapidly do lap times
deteriorate?'' or ``When does degradation become performance-limiting?''
In live racing, models must also be interpretable and computationally
efficient enough to inform real-time decisions. To address this, we
propose a Bayesian state-space model that represents tire degradation as
a latent process observed indirectly through lap times.

To the best of our knowledge, there are no examples in the literature
that apply state-space models to the phenomenon of tire degradation in
Formula 1. While prior research (e.g., Todd et al., 2025) has explored
deep learning approaches for tire energy prediction, those methods often
lack interpretability and explicit uncertainty quantification---features
that are crucial in operational race environments. Because of this, we
believe that state space models could be an asset to F1 teams looking to
gain an edge in predictive modeling.

Using publicly available data from the FastF1 Python API (Oehrly, 2025),
we analyze Lewis Hamilton's race at the 2025 Austrian Grand Prix. Lap
times are modeled as a function of fuel mass and latent tire pace, with
pit stops treated as state resets. Several extensions of the model are
tested, including compound-specific degradation rates and time-varying
degradation dynamics, with performance evaluated via cross-validation.
We use a Bayesian workflow for model development, visualization, and
comparison to allow for easier model implementation and parameter
interpretation.

The results show that the Bayesian state-space model fits lap-time data
well and outperforms an ARIMA(2,1,2) baseline in predictive accuracy. We
find limited evidence of statistically distinct degradation rates
between the hard and medium compounds in this case study--likely due to
a lack of data and modern tire management practices, where drivers
target consistent lap times to control wear. However, the model
nevertheless provides interpretable, real-time estimates of degradation
and uncertainty, offering a practical foundation for strategy modeling
even when compound effects are subtle.

Although presented for a single race, the framework generalizes
naturally to multi-race or multi-driver analyses, making it a promising
tool for quantifying degradation dynamics, evaluating strategies, and
improving model interpretability. In the remainder of this paper, we
will describe the dataset and processing steps taken to obtain the time
series of interest. Then, we will provide a brief background on
state-space models and describe the various models we propose for
analysis in this paper. Lastly, we perform model selection and
assessment, and discuss the results of fitting each model.

\section{Data}\label{data}

The FastF1 Python API (Oehrly, 2025) provides access to a vast array of
data pertaining to each F1 grand prix. Of interest to us are the
individual lap times for each driver, what tire compound the driver used
on each lap, and on what laps the driver pitted for a new set of tires.
Although there is much more available in the API, the aforementioned
items will be sufficient for the scope of this paper.

Figure 1 contains a plot illustrating the time series that we selected
for analysis in this paper. We chose Lewis Hamilton's race at the
Austrian Grand Prix because his race was fairly uneventful. No safety
car was deployed and he spent much of the race alone in free air,
unimpeded by slower vehicles. Data cleaning was minimal, consisting only
of the removal of laps in which the driver entered or exited the pit
lane, because such lap times would impart little to no information about
degradation.

Lastly, we include the amount of fuel in kilograms for the driver at
each lap as a covariate in our model. This data did not come from the
FastF1 Python API, and is assumed to start at 110 kilograms on lap 1 and
decay linearly to zero by the last lap. 110 kilograms is assumed to be
the starting value since that is the maximum allowed by regulation.
While a driver may start the race with less than 110 kilograms of fuel,
it is reasonable to assume that there is very little fuel left at the
end of the race because engineers try to optimize this as much as
possible to increase performance.

\begin{figure}

\centering{

\includegraphics{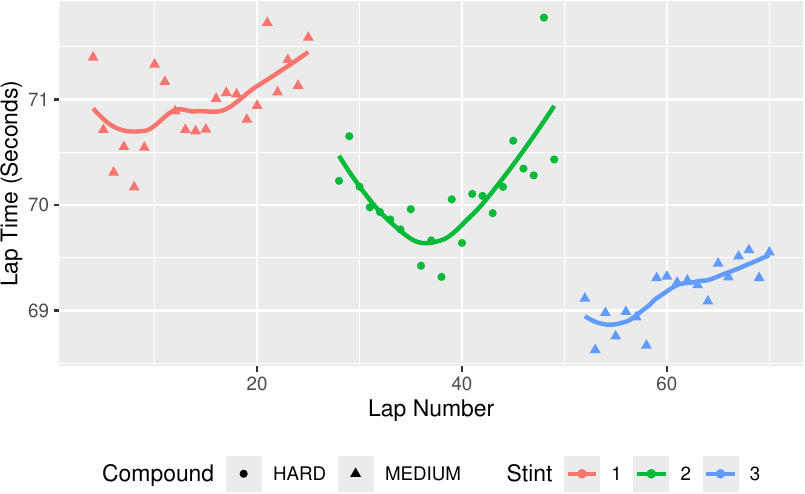}

}

\caption{\label{fig-myplot}Tire Degradation for Lewis Hamilton during
the Austrian Grand Prix. We can see a subtle but noticeable increase in
lap times throughout the course of each stint. In the second stint on
the hard tires, we can also see a warm-up period from laps 28 to 38.}

\end{figure}%

\section{Bayesian State-Space Model}\label{bayesian-state-space-model}

In this section we'll first briefly review state space models, then
describe in detail the process used to model latent degradation rates
through the observable lap time process. All models were fitted using
the software package Stan (Stan Development Team, 2020).

\subsection{Background}\label{background}

State-space models (SSMs) are a popular modeling framework for
time-series data due to their flexibility. They have found applications
in a wide range of areas, from ecological time series (Auger-Methe et
al., 2021), to financial data (Zeng and Wu, 2013), to sports analytics.
Within realm of sports analytics, Ridall et al.~(2025) used a Bayesian
state space model to predict Premier League football match outcomes by
assigning a latent state to the attacking and defending strengths of
each team. Duffield et al.~(2025) also introduced new approaches for
skill rating in competitive sports based on SSMs.

The defining feature of SSMs is their ability to model both a latent
unobserved time series via a state equation
\(\alpha_t \sim \pi(\cdot|\alpha_{t-1},y_{1:t-1})\), and an observation
time series \(Y_t | \alpha_t \sim f(\cdot|\alpha_t)\) that consists of
measurements which are related to the latent process. SSMs generally
make two assumptions:(1) the latent time series evolves as a (typically
first order) Markov Process and (2) the observations are independent of
one another when we condition on the latent states.

A variety of methods exist for estimation and inference, including the
Kalman filter for linear-Gaussian models (Kalman, 1960), Sequential
Monte Carlo methods for nonlinear or non-Gaussian systems, and particle
MCMC for joint inference on states and parameters (Andrieu, Doucet, and
Holenstein 2010). This paper uses Stan and MCMC (Auger-Méthé et al.,
2021) for model fitting and posterior sampling due to its flexibility
and ease of implementation.

\subsection{Base Model Specification and Parameter
Interpretation}\label{base-model-specification-and-parameter-interpretation}

We start with the observation equation for a driver's lap times:

\begin{align}
y_t &= \alpha_t + \gamma*fuel_t +\epsilon_t \\
\epsilon_t &\sim N(0,\sigma^2_\epsilon)
\end{align}

Here \(y_t\) represents the observed lap time for the driver on lap
\(t\). \(\alpha_t\) represents the true latent pace of the tires after
accounting for fuel and degradation. \(fuel_t\) is a covariate that
represents the derived amount of fuel in kilograms for the driver on lap
\(t\), and \(\gamma\) is the estimated increase in lap time due to an
additional kg of fuel. Lastly, \(\epsilon_t\) accounts for errors that
would result in a lap time being different from \(\alpha_t\) after
accounting for fuel loss. Possible sources of this error include driver
mistakes and the presence of other cars.

Now we present the process equation:

\begin{align}
\alpha_{t+1} &= (1-I_{pit_t})(\alpha_t+\nu) + I_{pit_t}(\alpha_{reset}) + \eta_t \\
\eta_t &\sim N(0,\sigma^2_\eta)
\end{align}

where:

\begin{align}
I_{pit_t} &= 
\begin{cases}
1 & \text{if driver has a new set of tires on lap t+1} \\
0 & \text{otherwise}
\end{cases} \\
t &\in \{1, 2, ..., T\}
\end{align}

As mentioned earlier, the latent states \(\alpha_t\) represent the true
pace (or lap time) that the tire is capable of. In the most basic
version of the model, we consider a linear rate of decay in lap times,
represented by the static parameter \(\nu\) in the model. However, we
allow for error in this decay process by including the term \(\eta_t\).
Perhaps most interesting in the process equation is the inclusion of an
indicator variable for pit stops. These allow us to reset the
degradation process to \(\alpha_{reset}\) after the driver puts on a new
set of tires, and then continue the degradation process as normal
afterwards.

\subsection{Extensions of the Basic
Model}\label{extensions-of-the-basic-model}

We will propose three extensions to this basic model. The first is to
estimate different degradation rates for each tire compound, and the
second is to allow the degradation rate \(\nu\) to increase over time.
The final extension will explore the benefits of using a skewed t
distribution to model the observation errors.

\subsubsection{Extension 1 - Compound Specific
Degradation}\label{extension-1---compound-specific-degradation}

As mentioned earlier, Formula 1 tires are designed to degrade at
different rates by Pirelli (Pirelli, n.d.). Therefore, a natural first
extension to make to the base model is to estimate different degradation
rates for each compound. With this in mind, our process equation
becomes:

\begin{align}
\alpha_{t+1} &= (1-I_{pit_t})(\alpha_t+\nu[compound_t]) + I_{pit_t}(\alpha_{reset}[compound_t]) + \eta_t \\
\eta_t &\sim N(0,\sigma^2_\eta)
\end{align}

where:

\begin{align}
compound_t &= 
\begin{cases}
1 & \text{if hard compound tires are used on lap t} \\
2 & \text{if medium compound tires are used on lap t} \\
3 & \text{if soft compound tires are used on lap t}
\end{cases}
\end{align}

As mentioned above, the main thrust of this extension is to estimate
different degradation rates and reset points for each tire compound used
by the driver.

\subsubsection{Extension 2 - Time-Varying
Degradation}\label{extension-2---time-varying-degradation}

When tires degrade, there is a loss of mechanical grip as rubber is torn
from the surface of the tire. One might well expect that this loss of
grip could lead to increased sliding and therefore a compounding of
degradation over time. With this in mind, we propose for the second
extension a model in which the degradation rate itself increases over
time. Under this extension, our process equations become:

\begin{align}
\alpha_{t+1} &= (1-I_{pit_t})(\alpha_t+\nu_t) + I_{pit_t}(\alpha_{reset}[compound_t]) + \eta_t \\
\nu_{t+1} &= (1-I_{pit_t})(\nu_t + \beta[compound_t]) + I_{pit_t}(v_{reset}) \\
\eta_t &\sim N(0,\sigma^2_\eta)
\end{align}

The most important difference here is that the degradation rate
\(\nu_t\) now changes with time. We estimate a parameter
\(\beta[compound_t]\) for each tire compound that represents an additive
increase to the degradation rate that occurs at each time step. It is
also important to note that since the degradation rate \(\nu_t\) is
allowed to vary with time, we must also include a reset parameter
\(\nu_{reset}\) so that the degradation rate can reset after a pit stop.

\subsubsection{Extension 3 - Skewed T
Distribution}\label{extension-3---skewed-t-distribution}

Our final extension to the base model is to use a skewed t distribution
(Hansen, 1994) for the observation error. Since drivers are given target
lap times by their engineers throughout the race, we expect to observe
extreme values predominately in the positive direction. For instance, a
driver might make a mistake that could lead to an increase of several
tenths of a second in lap time, but then return to the target times
given by the team. A positively skewed t distribution would capture the
possibility of extreme values in the positive direction. The base model
would have the same process equation, but the observation equation
becomes:

\begin{align}
y_t &= \alpha_t + \gamma*fuel_t +\epsilon_t \\
\epsilon_t &\sim Skewed-T(0,\sigma^2_\epsilon,\lambda, 2)
\end{align}

where zero is the mean, \(\sigma^2_\epsilon\) is again the variance,
\(\lambda\) is a skewness parameters that ranges from \(-1\) to \(1\),
and 2 is the degrees of freedom.

Because of the skewed t distribution's heavy tails (with lower degrees
of freedom), this model should be more robust to outliers than than
those using normally distributed errors.

\subsection{Discussion of Priors}\label{discussion-of-priors}

In general, we lean on moderately strong priors since we are relatively
data poor and have ample information to inform priors. Further,
informative priors improve convergence stability and speed, making the
model more practical in race conditions.

It should also be noted that lap times often differ by mere tenths of a
second. Therefore, priors which at first glance appear very strong, are
only moderately so. Given lap times vary by \textasciitilde0.5s per
stint, priors with SD = 0.1 represent plausible but informative
uncertainty levels.

\subsubsection{Base Model}\label{base-model}

The priors for our base model are:

\begin{align}
\sigma_\epsilon &\sim N^+(.3,.1^2) \\
\sigma_\eta &\sim N^+(.1,.1^2) \\
\nu &\sim N^+(.05,.1^2) \\
\alpha_{reset} &\sim N(69,.1^2)
\end{align}

We use a relatively strong prior on the observation standard errors
\(\sigma_\epsilon\) and \(\sigma_\eta\) because the degradation process
should have less error than the observation process. The observation
process can be affected by driver inconsistencies. Meanwhile the
underlying degradation process should remain relatively consistent
throughout.

We also use a half-normal prior on the degradation rate to restrict it
to be positive, as a negative overall degradation rate would be
nonsensical (if the degradation rate is not allowed to change with time
as in extension 3). We centered the prior at \(.05\) since the scale of
the data indicates the degradation rates will be small, but we still
believe that the degradation rate should be greater than zero.

Lastly, the prior on the \(\alpha_{reset}\) parameters are decided by
the long runs done during the free practice sessions. Before the race,
Lewis Hamilton's teammate Charles Leclerc drove a long-run on medium
tires suggesting that the race pace of the medium tires would be roughly
69-69.5 seconds per lap. As such, we centered the \(\alpha_{reset}\)
prior on 69.

\subsubsection{Extension 1 - Compound Specific
Degradation}\label{extension-1---compound-specific-degradation-1}

The error standard deviation priors for the compound specific
degradation model remain the same as before. However, the degradation
rate and state resets change since we have to estimate parameters for
each tire compound. We have the following extra priors in place of the
\(\nu\) and \(\alpha_{reset}\) of before:

\begin{align}
\nu[1] &\sim N^+(.01,.1^2) \\
\nu[2] &\sim N^+(.03,.1^2) \\
\nu[3] &\sim N^+(.05,.1^2) \\
\alpha_{reset}[1] &\sim N(69.5,.1^2) \\
\alpha_{reset}[2] &\sim N(69,.1^2) \\
\alpha_{reset}[3] &\sim N(68.5,.1^2)
\end{align}

Here \(\nu[1]\) is the degradation rate for the hard tire, \(\nu[2]\) is
the degradation rate for the medium tire, and \(\nu[3]\) is the
degradation rate for the soft tire. A similar pattern follows for the
reset parameters. Here, the priors for the degradation rates and resets
reflect our prior beliefs that harder tires should degrade more slowly,
but also start out slower. In contrast, the soft tires will start out
faster but degrade more quickly.

As mentioned in the previous section, there is data from the second free
practice session of that race weekend which suggested that the race pace
of the medium tires would be 69-69.5 seconds per lap. We use the lower
end of this spectrum to account for greater incentive to do faster lap
times during the actual race. Then, we make the hard tire reset value a
half second slower--and the soft tire a half second faster relative to
the medium tires--to reflect our beliefs that the soft tires will start
out faster and the hard tires will start out slower.

\subsubsection{Extension 2 - Time-Varying
Degradation}\label{extension-2---time-varying-degradation-1}

Here, the error standard deviation priors are the same as the base
model, and the reset parameters are the same as for the compound
specific degradation model. The main difference is that the degradation
priors are now on \(\beta\), and we include a prior for the degradation
state reset \(\nu_{reset}\).

\begin{align}
\beta[1] &\sim N^+(.005,.1^2) \\
\beta[2] &\sim N^+(.01,.1^2) \\
\beta[3] &\sim N^+(.02,.1^2) \\
\nu_{reset} &\sim N(0,.1^2)
\end{align}

\subsubsection{Extension 3 - Skewed T
Distribution}\label{extension-3---skewed-t-distribution-1}

For the final extension, we have only updated the observation equation.
We let \(\sigma_\epsilon\) have the same prior as in the base model and
put the following prior on the skew parameter \(\lambda\):

\[
\lambda \sim N(.5,.1^2)
\]

This prior reflects our belief that the distribution is skewed
positively. In fitting the model we used a parameterization of the
skewed t distribution in which \(\lambda\)--the skewness parameter--can
only take on values between \(-1\) and \(1\). This is reflected in the
parameter bounds of the Stan code used to fit the model.

Finally, we do not use a prior on the degrees of freedom because we know
that outliers can occur due to driver mistakes or getting stuck behind a
slower car, and therefore there is a need for heavy tails. Furthermore,
we want the model to fit quickly enough that it can provide strategic
information during a race. Adding a prior on the degrees of freedom
would make the model take longer to run with little added benefit.

\section{Results}\label{results}

In this section we will discuss the results of fitting the various
models. In particular, we will discuss estimates of degradation rates
across tire compound and prediction of lap times.

\subsection{Model Selection}\label{model-selection}

We used rolling-origin-recalibration cross validation to perform model
selection (Tashman 2000). We describe the cross validation scheme below.
Let \(S_i\) represent the last lap of stint \(i\), and let
\(i \in \{1, ...,N\}\) where \(N\) is the number of stints in the race.
Lastly, note that \(\lceil x \rceil\) represents the ceiling function
for some \(x\in\mathbb{R}\). We used the following cross validation
scheme:

\textbf{Stint 1}

Fold 1 - Train: \([1, 2, ..., \lceil\frac{3}{4}S_1\rceil]\) Test:
\([\lceil\frac{3}{4}S_1\rceil + 1]\)

Fold 2 - Train:
\([1, 2, ..., \lceil\frac{3}{4}S_1\rceil, \lceil\frac{3}{4}S_1\rceil + 1]\)
Test: \([\lceil\frac{3}{4}S_1\rceil + 2]\)

\ldots{}

Fold \(\frac{S_1}{4}\) - Train: \([1, 2, ..., S_1-1]\) Test: \([S_1]\)

\ldots{}

\textbf{Stint N}

Fold 1 - Train: \([1, 2, ..., \lceil\frac{3}{4}S_N\rceil]\) Test:
\([\lceil\frac{3}{4}S_N\rceil + 1]\)

Fold 2 - Train:
\([1, 2, ..., \lceil\frac{3}{4}S_N\rceil, \lceil\frac{3}{4}S_N\rceil + 1]\)
Test: \([\lceil\frac{3}{4}S_N\rceil + 2]\)

\ldots{}

Fold \(\frac{S_N}{4}\) - Train: \([1, 2, ..., S_N-1]\) Test: \([S_N]\)

In this way we perform cross validation on each stint of the driver's
race, and calculate the root mean squared predictive error for each
stint so that we can analyze model performance at the stint-level.
Letting \(\hat{y}_j\) denote our prediction for the \(jth\) lap, our
test statistic is then:

\[
RMSPE_i = \sqrt{\frac{1}{S_i-\lceil\frac{3}{4}S_i\rceil}\sum_{j=\lceil\frac{3}{4}S_i\rceil}^{S_i}(y_j - \hat{y_j})^2}
\]

We performed cross validation for the base model described above and the
three extensions. In addition, we include an ARIMA(2,1,2) model as a
baseline. The first-order differencing removes the non-stationary
degradation trend, while the AR(2) and MA(2) components capture
short-term autocorrelation and transient shocks in lap times. This
specification provides a flexible classical time-series benchmark
against which to evaluate the state-space models.'' Results can be seen
in Table 1.

\begin{longtable}[]{@{}
  >{\raggedright\arraybackslash}p{(\columnwidth - 10\tabcolsep) * \real{0.1127}}
  >{\raggedleft\arraybackslash}p{(\columnwidth - 10\tabcolsep) * \real{0.1831}}
  >{\raggedleft\arraybackslash}p{(\columnwidth - 10\tabcolsep) * \real{0.1549}}
  >{\raggedleft\arraybackslash}p{(\columnwidth - 10\tabcolsep) * \real{0.1690}}
  >{\raggedleft\arraybackslash}p{(\columnwidth - 10\tabcolsep) * \real{0.1690}}
  >{\raggedleft\arraybackslash}p{(\columnwidth - 10\tabcolsep) * \real{0.2113}}@{}}
\caption{Cross Validation Results - RMSPE}\tabularnewline
\toprule\noalign{}
\begin{minipage}[b]{\linewidth}\raggedright
\end{minipage} & \begin{minipage}[b]{\linewidth}\raggedleft
ARIMA(2,1,2)
\end{minipage} & \begin{minipage}[b]{\linewidth}\raggedleft
Base Model
\end{minipage} & \begin{minipage}[b]{\linewidth}\raggedleft
Extension 1
\end{minipage} & \begin{minipage}[b]{\linewidth}\raggedleft
Extension 2
\end{minipage} & \begin{minipage}[b]{\linewidth}\raggedleft
Skewed T Dist.
\end{minipage} \\
\midrule\noalign{}
\endfirsthead
\toprule\noalign{}
\begin{minipage}[b]{\linewidth}\raggedright
\end{minipage} & \begin{minipage}[b]{\linewidth}\raggedleft
ARIMA(2,1,2)
\end{minipage} & \begin{minipage}[b]{\linewidth}\raggedleft
Base Model
\end{minipage} & \begin{minipage}[b]{\linewidth}\raggedleft
Extension 1
\end{minipage} & \begin{minipage}[b]{\linewidth}\raggedleft
Extension 2
\end{minipage} & \begin{minipage}[b]{\linewidth}\raggedleft
Skewed T Dist.
\end{minipage} \\
\midrule\noalign{}
\endhead
\bottomrule\noalign{}
\endlastfoot
Stint 1 & 0.613 & 0.358 & 0.355 & 0.386 & 0.325 \\
Stint 2 & 0.727 & 0.673 & 0.692 & 0.670 & 0.601 \\
Stint 3 & 0.180 & 0.139 & 0.140 & 0.163 & 0.156 \\
Total & 1.520 & 1.169 & 1.187 & 1.218 & 1.082 \\
\end{longtable}

Since we obtained samples from the one step ahead predictive
distributions using Stan, we also use the Continuous Rank Probability
Score (Matheson and Winkler, 1976) with the same cross validation scheme
as above to evaluate our probabilistic forecasts. Let \(CRPS_{i,j}\)
denote the CRPS for the \(jth\) lap in the \(ith\) stint. The overall
CRPS for the stint is:

\[
CRPS_i = \frac{\sum_{j=\lceil\frac{3}{4}S_i\rceil}^{S_i}CRPS_{i,j}}{S_i-\lceil\frac{3}{4}S_i\rceil}
\]

In other words, we take the average of each CRPS within a stint to get a
stint-level CRPS. Similarly, the overall \(\bar{CRPS}\) is an average of
each stint-level CRPS. The results can be seen in Table 2. Note that a
smaller CRPS is indicative of a better forecast.

\newpage

\begin{longtable}[]{@{}lrrrrr@{}}
\caption{Continuous Rank Probability Score}\tabularnewline
\toprule\noalign{}
& Base & Extension 1 & Extension 2 & Skew-T & ARIMA(2,1,2) \\
\midrule\noalign{}
\endfirsthead
\toprule\noalign{}
& Base & Extension 1 & Extension 2 & Skew-T & ARIMA(2,1,2) \\
\midrule\noalign{}
\endhead
\bottomrule\noalign{}
\endlastfoot
Stint 1 & 0.201 & 0.201 & 0.220 & 0.184 & 0.424 \\
Stint 2 & 0.377 & 0.391 & 0.396 & 0.316 & 0.411 \\
Stint 3 & 0.112 & 0.115 & 0.119 & 0.106 & 0.137 \\
\(\bar{CRPS}\) & 0.230 & 0.236 & 0.245 & 0.202 & 0.324 \\
\end{longtable}

The SSM with skewed t errors is shown to be the best in terms of RMSPE,
beating the base model by nearly a tenth. Given the scale of the data,
this indicates that the skewed t model performs best on out-of-sample
data. Interestingly, this model performs much better than the others in
the second stint where there is an extreme outlier in the positive
direction. Since performance of the state space models is close among
those with normal errors we will still examine them all in the remaining
sections, but for out-of-sample prediction we deem the SSM with skewed t
errors to perform best.

We see a similar pattern in the Continuous Rank Probability Scores
(CRPS) for each of the models (Table 2). The skewed t model beats the
other models in all stints, but again performs particularly well in
stint 2. The CRPS takes into account the full forecast distribution, so
forecasts for the skewed t model are likely able to capture extreme
values that the normally distributed models are unable to.

\subsection{Model Assessment}\label{model-assessment}

We obtained posterior samples for the state-space models via Hamiltonian
Monte Carlo sampling with 4 chains of 15000 samples each after 15000
burn-in iterations. R-hat values for all parameters were less than 1.01,
indicating adequate convergence.

It can be seen from Figure 2 that the models all fit the data reasonably
well, and are fairly similar. Notably however, the skewed t distribution
is not nearly as affected by outliers in the first and second stints,
leading to a better fit.

\begin{figure}[H]

{\centering \includegraphics[width=1\textwidth,height=1\textheight]{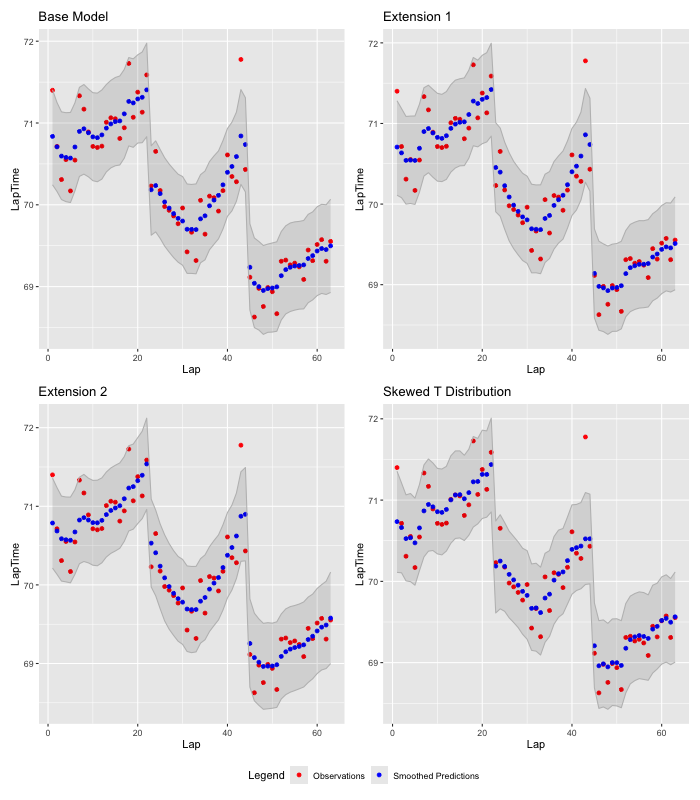}

}

\caption{Fit of the various models with 90\% credible intervals. The
smoothed predictions are based on the entire time series, as opposed to
one-step-ahead predictions which are based solely on observations that
occur before the prediction. The smoothed predictions are a basic check
that show the model fits the data well. Interestingly, we can see that
the skewed t model is not nearly as affected by the outlier on lap 43.}

\end{figure}%

\subsection{Fitting the First and Second Extensions of the Base
Model}\label{fitting-the-first-and-second-extensions-of-the-base-model}

In the previous section we saw that the skewed t model performed best.
While we did expect this model to perform better than the base model, it
is surprising that the compound-specific and time-varying degradation
models were outperformed by the base model, especially considering that
the entire purpose of having different tire compounds in F1 is so that
certain compounds will degrade more quickly.

\begin{longtable}[]{@{}lrrr@{}}
\caption{Estimated Degradation Rates with 95\% Credible
Intervals}\tabularnewline
\toprule\noalign{}
& \(\nu\) & 2.5\% & 97.5\% \\
\midrule\noalign{}
\endfirsthead
\toprule\noalign{}
& \(\nu\) & 2.5\% & 97.5\% \\
\midrule\noalign{}
\endhead
\bottomrule\noalign{}
\endlastfoot
Hard & 0.054 & 0.004 & 0.133 \\
Medium & 0.060 & 0.009 & 0.120 \\
\end{longtable}

Table 3 shows the estimated values of \(\nu\) for the first extension to
the base model. These estimates are for the hard and medium tires used
by Lewis Hamilton in the Austrian Grand Prix, along with bounds for a
95\% credible interval. Based on this model, we estimate that Lewis
Hamilton loses 5.4 hundredths of a second per lap and 6 hundredths of a
second per lap due to tire degradation for hard and medium compound
tires respectively, with large uncertainty.

While we do estimate a slightly higher degradation rate for the medium
compound tires, the credible intervals have a large degree of overlap,
indicating that the data provides little evidence that there is a
difference in degradation rate between the compounds.

\begin{longtable}[]{@{}lrrr@{}}
\caption{Estimated \(\beta\) with 95\% Credible
Intervals}\tabularnewline
\toprule\noalign{}
& \(\beta\) & 2.5\% & 97.5\% \\
\midrule\noalign{}
\endfirsthead
\toprule\noalign{}
& \(\beta\) & 2.5\% & 97.5\% \\
\midrule\noalign{}
\endhead
\bottomrule\noalign{}
\endlastfoot
Hard & 0.011 & 0.003 & 0.020 \\
Medium & 0.010 & 0.002 & 0.018 \\
\end{longtable}

\begin{figure}

\centering{

\includegraphics{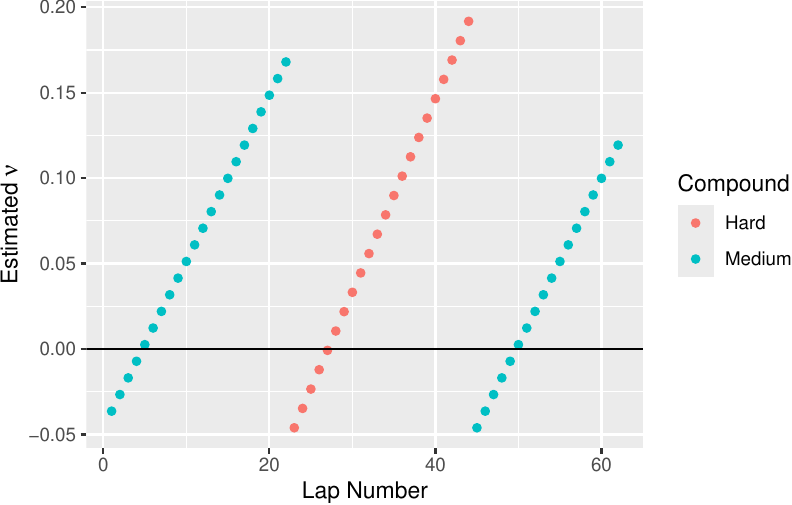}

}

\caption{\label{fig-myplot2}Degradation rates for each lap as estimated
by the time-varying degradation model. Interestingly, the degradation
rate begins below zero in each stint, indicating the model is capable of
capturing a warm-up period for the tires before they begin degrading.}

\end{figure}%

From Table 4 it can be seen that the estimated \(\beta\) of the second
extension to the base model for the hard tire compound is slightly
greater than that for the medium tire compound. Once again however, the
credible intervals show a large degree of overlap, indicating little
evidence based on the data that the two parameters are different.

Interestingly, the degradation rate \(\nu\) starts negative, then
increases to roughly .175 before a pit stop (see figure 3). This makes
sense given that the tires require a warm up period before reaching
their optimum operating window(Kelly and Sharp, 2012). Thus, while the
skewed t model performed best in terms of predictive accuracy, we can
still glean interesting insights from the more complicated models.

Of course, in both model extensions we see that our degradation
estimates do not meaningfully change across the tire compounds used.
This is likely why the base model performs better in terms of prediction
than our extensions. Another important consideration is that drivers
strive to achieve target lap times set by their engineers during a race.
Thus, they are not driving at the absolute limit and are actively trying
to manage their degradation rates. This partly explains why we tend to
see a linear decay. That being said, each stint only has around 20 laps,
so we don't have much data to differentiate what would likely be a small
effect size.

\subsection{Prediction of Lap Times with
Uncertainty}\label{prediction-of-lap-times-with-uncertainty}

One benefit of these models is the ability to quickly assimilate new
data points and get predictions for the next lap time with uncertainty
intervals. For example, if we run the base model on laps 1-43 to predict
lap 44, we get the results seen in figure 4.

Figure 5 also supports our claim that the models do a good job at
forecasting the next lap time. It takes between 15 and 30 seconds to run
the base and extension 1 models, giving plenty of time to use the
results for decision making in the rest of a lap. In addition, if the
fully extended model with increasing degradation rate \(\nu\) is used, a
team could use a certain threshold of \(\nu\) beyond which they consider
a pit stop. In other words, when the degradation rate gets too high,
teams can begin seeking an optimal window in which the driver can be
pitted. Furthermore, teams could use these results to compare strategies
and determine the effects of pitting in optimizing their overall race
time.

\begin{figure}

\centering{

\includegraphics{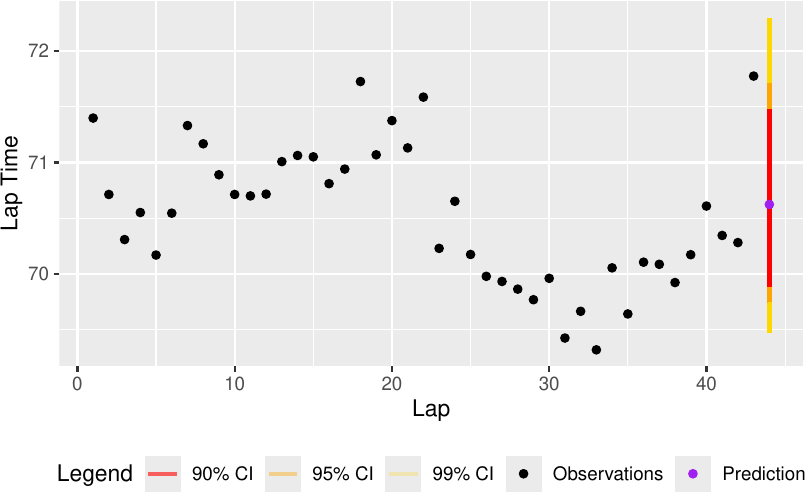}

}

\caption{\label{fig-myplot3}One-step-ahead prediction of lap 44, given
laps 1 to 43. Our point estimate is clearly robust to the outlier on the
previous lap. We can also see evidence of the skewed t observation
errors in the credible intervals. While the 90\% interval appears fairly
symmetric, we see that increasing the probability extends the interval
farther in the positive direction than in the negative direction
(relative to the point estimate).}

\end{figure}%

\clearpage

\begin{figure}[H]

{\centering \includegraphics[width=1\textwidth,height=1\textheight]{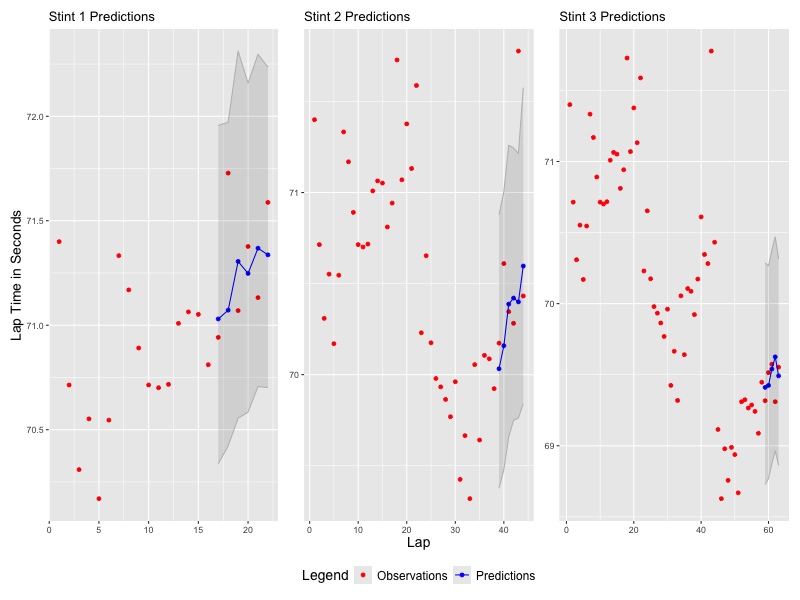}

}

\caption{One step ahead predictions with 90\% credible intervals for the
skew t model. Generally speaking, the models do a good job of predicting
the next lap time. The uncertainty intervals almost always contain the
observation.}

\end{figure}%

\subsection{Limitations and
Considerations}\label{limitations-and-considerations}

Firstly, the Austrian Grand Prix did not have a safety car. Safety cars
come onto the track when there has been a serious crash, and all drivers
are forced behind the safety car to limit their speeds. As such, driving
under safety car conditions drastically reduces degradation since the
drivers are limited to much slower speeds. Such a situation could be
easily accounted for by extending our model to suspend the degradation
process for laps done under safety car conditions.

Secondly, Lewis Hamilton's drive at the Austrian Grand Prix was fairly
uneventful and so he was minimally impeded by the drivers ahead. If a
driver gets stuck behind a slower car, this can cause an
\emph{artificial} increase in lap times that isn't due to tire
degradation. The easiest way to address this if necessary is to add a
covariate to the observation equation for the distance to the driver
ahead. Future work, however, will likely look into more sophisticated
ways to address this, such as a multivariate time series with all
drivers and dependent errors based on the distance to the driver ahead.

\section{Conclusion}\label{conclusion}

This paper introduced a Bayesian state-space framework for modeling tire
degradation in Formula 1 racing, demonstrating that such models can
capture the latent deterioration of tire performance while providing
interpretable and probabilistic predictions of lap times. Using Lewis
Hamilton's 2025 Austrian Grand Prix as a case study, the proposed
approach showed superior predictive performance relative to an
ARIMA(2,1,2) baseline, particularly when observation errors were modeled
with a skewed t distribution to account for asymmetric driver mistakes.

Although degradation rates between tire compounds were not found to
differ greatly, teams that have access to more telemetry data could
likely discern meaningful differences between tire compounds. The
state-space framework's ability to assimilate new data in real time and
output predictive uncertainty makes it a strong candidate for
integration into race strategy tools.

Future work should extend the model across multiple races and drivers to
better quantify compound-specific degradation patterns, refine priors
using telemetry or surface-temperature data, and explore hierarchical
structures for team- or track-level effects. Overall, the Bayesian
state-space approach provides a statistically principled and
computationally efficient foundation for studying tire behavior and
optimizing strategy in Formula 1.

\section{References}\label{references}

Andrieu C, Doucet A and Holenstein R (2010) Particle Markov Chain Monte
Carlo Methods. Journal of the Royal Statistical Society Series B
Statistical Methodology 72(3): 269--342.

Auger-Méthé M, Newman K, Cole D, Empacher F, Gryba R, King AA,
Leos-Barajas V, Mills Flemming J, Nielsen A, Petris G and Thomas L
(2021) A guide to state--space modeling of ecological time series.
Ecological Monographs 91(4): e01470.

Duffield S, Power S and Rimella L (2024) A state-space perspective on
modelling and inference for online skill rating. Journal of the Royal
Statistical Society Series C Applied Statistics 73(5): 1262--1282.

Farroni F, Sakhnevych A and Timpone F (2016) Physical modelling of tire
wear for the analysis of the influence of thermal and frictional effects
on vehicle performance. Proceedings of the Institution of Mechanical
Engineers Part L Journal of Materials Design and Applications 231(1-2):
151--161.

Hansen BE (1994) Autoregressive Conditional Density Estimation.
International Economic Review 35: 705--730.

Kalman RE (1960) A new approach to linear filtering and prediction
problems. Journal of Engineering for Industry 82(1): 35--45.

Kelly DP and Sharp RS (2012) Time-optimal control of the race car:
influence of a thermodynamic tyre model. Vehicle System Dynamics 50(4):
641--662.

Matheson JE and Winkler RL (1976) Scoring Rules for Continuous
Probability Distributions. Management Science 22(10): 1087--1096.

Oehrly T (2025) FastF1: A Python package for accessing and analyzing
Formula 1 results, schedules, timing data, and telemetry (version
3.6.1). Available at: https://github.com/theOehrly/Fast-F1 (accessed 6
November 2025).

Pirelli (n.d.) F1 Tires: Details and Technical Data. Available at:
https://www.pirelli.com/tires/en-us/motorsport/f1/tires (accessed 6
November 2025).

Giles R (2024) Charles Leclerc wins Italian F1 GP for Ferrari after
one-stop gamble. The Guardian, 1 September.

Ridall PG, Titman AC and Pettitt AN (2025) Bayesian state-space models
for the modelling and prediction of the results of English Premier
League football. Journal of the Royal Statistical Society Series C
Applied Statistics 74(3): 717--742.

Stan Development Team (2020) Stan: A Probabilistic Programming Language.
Journal of Statistical Software 95(1): 1--29.

Tashman L (2000) Out-of-sample tests of forecasting accuracy: An
analysis and review. International Journal of Forecasting 16(4):
437--450.

Todd J et al.~(2025) Explainable Time Series Prediction of Tyre Energy
in Formula One Race Strategy. arXiv preprint arXiv:2501.04067.

Zeng Y and Wu S (eds) (2013) State-Space Models: Applications in
Economics and Finance. New York: Springer.

\end{document}